\documentclass[conference]{IEEEtran}

\ifCLASSINFOpdf

\else

\fi

\usepackage{amsmath}
\usepackage{amssymb}
\usepackage{lineno}
\usepackage{url}
\usepackage{enumerate}
\usepackage{xcolor}
\newcommand{\BfPara}[1]{{\noindent\bf#1.} \hspace{1mm}}
\usepackage{xspace}
\newcommand{\etal}{{\em et al.}\xspace}

\newcommand{\ie}{{\em i.e.,}\xspace}

\usepackage{xspace}
\usepackage{setspace}

\usepackage{makeidx} 
\usepackage{tikz}
\usetikzlibrary{arrows}
\usepackage[inline]{enumitem}

\begin{document}
\title{Domain Name System Security and Privacy: \\ Old Problems and New Challenges} 

\author{\IEEEauthorblockN{Ah Reum Kang}
\IEEEauthorblockA{University at Buffalo\\
Buffalo, NY, USA}
\and
\IEEEauthorblockN{Jeffrey Spaulding}
\IEEEauthorblockA{University at Buffalo\\
Buffalo, NY, USA}
\and
\IEEEauthorblockN{Aziz Mohaisen}
\IEEEauthorblockA{University at Buffalo\\
Buffalo, NY, USA}
}


%


\maketitle

\begin{abstract}
The domain name system (DNS) is an important protocol in today's Internet operation, and is the standard naming convention between domain names, names that are easy to read, understand, and remember by humans, to IP address of Internet resources. The wealth of research activities on DNS in general and security and privacy in particular suggest that all problems in this domain are solved. Reality however is that despite the large body of literature on various aspects of DNS, there are still many challenges that need to be addressed. In this paper, we review the various activities in the research community on DNS operation, security, and privacy, and outline various challenges and open research directions that need to be tackled. 
\end{abstract}

%
\IEEEpeerreviewmaketitle

\section{Introduction}
Since the inception of the domain name system (DNS) in 1983~\cite{mockapetris1987domain}, there has been a large body of work on understanding its operation, security, and privacy. Issues as understanding the DNS ecosystem\cite{AppelbaumGWW12,SSAC10,rfc5395,MockapertisD88}, resolvers behavior~\cite{SchompRA16,GaoYCPGJD13,CallahanAR13}, security issues of resolution\cite{GoldbergNPRVZ15,ShulmanW14,HerzbergS13,AtenieseM01,Conrad01}, applications for malicious actors detection and profiling\cite{PaxsonCJRSSSTVW13,AntonakakisPLVD11,WeaverKP11,AntonakakisPDLF10,KrishnanM10,ChangMWC15,WangMCC15}, DNS privacy\cite{ZhuHHWMS15,Bortzmeyer13a,MohaisenM15,Shulman14,Bortzmeyer13,KrishnanM10,Castillo-PerezG09}, etc. have been widely researched.  


The large body of literature on DNS operation, security, and privacy suggests that the area is mature, and problems are well understood. However, reality is contrary to this suggestion. Recently, adversaries have become very ``creative'' about the way they use DNS for launching attacks, moving from simple to sophisticated usage~\cite{HerzbergS14,Rossow14}. The DNS ecosystem has evolved to include many players, such as open DNS resolvers, which include trusted, untrusted, and semi-trusted ones, making it very difficult to reason about its resolution and operation. The rise of nation-state adversaries, with their unique capabilities compared to typical adversaries (ISP-level of an individual malicious actor) call for the further understanding of how they can affect the operation of the Internet in general, and DNS in particular~\cite{BarnesSJHTTHB15}. 


Despite the large body of work on DNS, the rise of new attacks suggests that DNS operation, security, and privacy are still one of the significant areas to explore with various issues to address. Those issues are not necessarily new, like addressing pervasive adversaries, privacy, and new forms of attacks, but could be issues to do with problems already explored in the past: as the behavior of DNS users (benign and malicious) evolves over time, this calls for further exploration to incorporate such behavior in characterizing, identifying, and detecting misuse. As new entities and operational realities and functions get incorporated in the DNS system, their role and how they affect end-to-end guarantees, and services built on top of DNS, need to be understood. 

Believing in the important role that it plays today and will play the future, we set out to present the current opportunities and challenges of DNS. We summarize all of the major thrusts of research on the topic and explore some of our ongoing research activities, highlighting some challenges, and call on the community to help address them.

\BfPara{Organization} In \ref{sec:research} we introduce the research opportunities. In \ref{sec:challenges} we introduce various challenges and directions. In \ref{sec:concluding} we introduce concluding remarks. 

\section{A Review of the Research Opportunities}\label{sec:research}
We review various avenues of research uncovered in the rich literature on DNS. The main objective of this review is to highlight DNS security and privacy, with the secondary objective being the operation of DNS. We focus on research for understanding the DNS ecosystem (\ref{sec:ecosystem}), DNS security (\ref{sec:security}) and DNS privacy (\ref{sec:privacy}).

\vspace{-2mm}
\subsection{DNS Economics and System Analysis}\label{sec:ecosystem}
The role that each entity plays, including clients and resolvers, and how these roles interact is the determinant factor for understanding the DNS ecosystem and its various pieces of complex infrastructure. In the following, we review two crucial areas that have been explored on that front which includes understanding the behavior of DNS and DNS blocking.

\BfPara{Understanding DNS Behavior} Callahan \etal~\cite{CallahanAR13} passively monitored DNS and related traffic within a local network to understand server behavior. Shcomp \etal~\cite{SchompRA16} presented a characterization of DNS clients for developing an analytical model of client interactions with the larger DNS ecosystem. Banse \etal~\cite{BanseHF12} studied the feasibility of behavior-based tracking in a real-world setting. 
Schomp \etal~\cite{SchompCRA13} presented methodologies for efficiently discovering the complex client-side DNS infrastructure. They further developed measurement techniques for isolating the behavior of the distinct actors in the infrastructure. Shulman and Waidner~\cite{ShulmanW15} explored name servers that use server-side caching, characterized the operators of the server-side caching resolvers and their motivations. Hao \etal~\cite{HaoFP11} explored the behavioral properties of domains using the DNS infrastructure associated with the domain and the DNS lookup patterns from networks who are looking up the domains initially. Behavior-based tracking is a threat, allowing attackers to track users passively. Multiple sessions of a user are linked exploiting characteristic patterns gathered from network traffic. For user's privacy, daily changing IP addresses offer limited protection against behavior-based tracking. Thus, lightweight methods that help to prevent profiling and tracking users without their consent are needed.

\BfPara{DNS Blocking} 
Thomas \etal~\cite{ThomasLS14} examined the NXD (Non-eXistent Domain) request patterns observed at the root and recursive name servers to gauge the effectiveness of collision blocking techniques. Scaife \etal~\cite{ScaifeCT15} presented an anonymous domain registrar. Appelbaum and Muffett~\cite{AppelbaumM15} proposed blocking special queries (\ie .onion) to improve Tor's privacy. DNS blocking has a limitation. In order to ensure DNS privacy, blocking should be implemented at once. However, it is almost impossible for all browsers and recursive resolvers to perform blocking simultaneously. Understanding how blocking may affect users who are not performing blocking but are sharing the same DNS infrastructure is required.

\vspace{-2mm}
\subsection{DNS Security}\label{sec:security}
DNS security is one of the well-explored areas in the literature, where work has been focused on analyzing and detecting DNS vulnerabilities and malicious domains. We review some of the outstanding work on each of these topics.

\BfPara{DNS Vulnerability} Schomp \etal~\cite{SchompCRA14} measured vulnerability to DNS record injection attacks and found that record injection vulnerabilities are fairly common even years after some of them were first uncovered. Dagon \etal~\cite{DagonPLL08} documented how attackers are using ``rogue'' DNS servers to create malicious DNS resolution paths, showing dozens of viruses that corrupt resolution paths and noting hundreds of URLs discovered per week that performed drive-by alterations of host DNS settings. Xu \etal~\cite{XuBSY13} quantitatively analyzed several techniques for hiding malicious DNS activities. Jackson \etal~\cite{JacksonBBSB09} evaluated the cost-effectiveness of mounting DNS rebinding attacks. Schomp \etal~\cite{SchompAR14} addressed vulnerabilities in shared DNS resolvers by removing them entirely and leaving recursive resolution to clients, showing that the cost of this approach is modest and arguing that it strengthens the DNS by reducing the attack surface. Dagon \etal~\cite{DagonAVJL08} proposed a technique to make DNS queries more resistant to poisoning attacks. Chen \etal \cite{ChenMP15} proposed a lightweight DNS record's TTL optimization for consistency. Despite such work, extending the literature for user-intention-based anomaly detection method to identify anomalous DNS traffic is an open challenge.

\BfPara{Detecting Malicious Domain Names} Various works have been proposed for detecting malicious domain names using DNS behavioral profiling~\cite{AntonakakisPLVD11,AntonakakisPNVALD12,RahbariniaPA15,JonesFPWA16,HaoFP11,BilgeKKB11,LuoTZSLNM15,PerdisciCG12,DagonPLL08}. For example, Antonakakis \etal~\cite{AntonakakisPLVD11}  proposed a novel detection system called \textit{Kopis} for detecting malware-related domain names by passively monitoring DNS traffic at the upper levels of the DNS hierarchy. Johnes \etal~\cite{JonesFPWA16} presented techniques for detecting unauthorized DNS root servers on the Internet using primarily endpoint-based measurements.  Yadav \etal~\cite{YadavRRR10} developed a method to detect domain fluxes in DNS traffic by looking for patterns inherent to domain names that are generated algorithmically.  Antonakakis \etal~\cite{AntonakakisPDLF10} suggested a dynamic reputation system for DNS called \textit{Notos}, which uses passive DNS query data and analyzes the network and zone features of domains to indicate if a new domain is malicious or legitimate. Gao \etal \cite{GaoYCPGJD13} presented an innovative approach to detect previously unknown malicious domains by simply using temporal correlation in DNS queries. Szurdi \etal \cite{SzurdiKCSFK14} performed a comprehensive study of ``typosquatting'' (\ie \textit{the deliberate registration of domains containing typos}) within the \texttt{.com} TLD, showing typo domains identified by lexical analysis are truly typographical variants of their target domain names. Despite such measures, the integrity and availability of Internet communication rely on replies from the DNS root name servers. Thus, it is important to detect DNS root manipulation when it does occur, even though it is rare.

\BfPara{Modeling adversaries} There have been some works modeling DNSSEC (Domain Name System Security Extensions) adversaries, such as Bau and Mitchell~\cite{BauM10} who formally modeled the cryptographic operations in DNSSEC and discovered a forgery vulnerability.  Herzberg \etal~\cite{HerzbergS13} presented a comprehensive overview of challenges and  pitfalls of DNSSEC, including vulnerable configurations, interoperability of incremental deployment, and challenges due to  super-sized DNS responses. Although DNSSEC deployment is still very limited, it has already been abused in several of the largest DoS attacks. These attacks often deter domains from deploying DNSSEC. Goldberg \etal~\cite{GoldbergNPRVZ15} demonstrated that since current DNSSEC deployments with support for NSEC and/or NSEC3 are vulnerable to zone enumeration attacks, they proposed a new cryptographic construction called \textit{NSEC5}--which proved to be a secure DNSSEC denial-of-existence mechanism. DNSSEC does not protect against denial of service attacks. DNSSEC makes DNS vulnerable to a new type of denial of service attacks based on cryptographic operations against security-aware resolvers and name servers, as an attacker can attempt to use DNSSEC mechanisms to consume victim's resources. 

\subsection{DNS Privacy}\label{sec:privacy}
DNS privacy is quickly becoming one of the most emergent issues in the DNS research community. Despite the large body of the literature on this problem, including but not limited to: 
\begin{enumerate*}[label=(\roman*)]
	\item quantification of DNS privacy leakage
	\item designs to improve privacy
	\item DNS encryption as a vehicle to improve privacy, and
	\item various standard body activities, 
\end{enumerate*}
many in the academic research community are still doubtful about the privacy risks in DNS~\cite{priv}. Notwithstanding such doubts, we review prior work on DNS privacy and open directions.

\BfPara{Quantifying DNS Leakage} Konings \etal~\cite{KoningsBSW13} collected a one-week dataset of multicast domain name system (mDNS) announcements at a university and showed that queries and device names leak plenty of information about users. Krishnan \etal~\cite{KrishnanM10} demonstrated privacy leakage by prefetching and showed that it is possible to infer the likelihood of search terms issued by clients by analyzing the context obtained from the prefetching queries.  Zhao \etal~\cite{ZhaoHS07} analyzed the complete DNS query process and discussed privacy disclosure problems in each step: client-side, query transmission process, and DNS server-side. They proposed a privacy-preserving query scheme called ``Range Query", which decreases privacy disclosure in the whole DNS query process. Paxson \etal~\cite{PaxsonCJRSSSTVW13} developed a measurement procedure to limit the amount of information a domain can receive surreptitiously through DNS queries to an upper bound specified by a security policy, with the exact setting representing a tradeoff between the scope of potential leakage versus the quantity of possible detections. Castillo-Perez and Garcia-Alfaro~\cite{Castillo-PerezG09} evaluated DNS privacy-preserving approaches, and pointed out the necessity of additional measures to enhance their security. When mobile devices are operated in public wireless networks, current implementations pose several privacy risks. The default naming practices of devices names need to be revised and users need to be able to limit service discovery to a selected set of networks.

\BfPara{Designs for Improving Privacy}
Due to the ubiquity of privacy risks, efforts are constantly being made in both academia and industry for preserving privacy in DNS communications. Zhao~\etal~\cite{ZhaoHS07} propose to ensure the DNS privacy by concealing the actual queries using noisy traffic. Noisy traffic increases of latency and bandwidth during the execution and resolution of queries. Castillo-Perez \etal~\cite{Castillo-PerezG09} evaluated this approach and demonstrated that the privacy ensured by added queries is somewhat difficult to analyze and that the technique introduces additional latency and overhead, making it less practical. An extended algorithm to ensure privacy, while improving performance, is also introduced by Castillo-Perez \etal~\cite{Castillo-PerezG09} which uses both noisy traffic and private information retrieval (PIR) techniques. They pointed out to serious security flaws on both proposals if active attackers can target those mechanisms. These flaws on an improved method of the two proposals still require additional improvements to be effective.

Techniques which employ certain flavors of encryption have also been studied. For example, Herrmann \etal~\cite{HerrmannFLF14} proposed a lightweight privacy-preserving implementation called \textit{EncDNS} which essentially replaces third-party resolvers and provides client software that forwards queries to it through conventional DNS forwarders. Since EncDNS provides an end-to-end encryption, forwarders will not know the contents of the queries. Lu and Tsudik~\cite{LuT10} proposed a Privacy-Preserving DNS (PPDNS) built on top of a distributed hash tables (DHTs) and computational PIR to obtain a reasonably high level of privacy for name resolution queries.

\BfPara{DNS Encryption} 
As mentioned previously, Herrmann \etal~\cite{HerrmannFLF14} presented a novel lightweight privacy-preserving name resolution service called EncDNS to serve as a replacement for conventional third-party resolvers. The EncDNS protocol, which is based on \textit{DNSCurve}, encapsulates encrypted messages in standards-compliant DNS messages. Zhu \etal~\cite{ZhuHHWMS15} proposed T-DNS to address privacy leakage problems using transport-layer security (TLS) to enable a user's privacy against their DNS resolvers and {\em optionally} the authoritative servers. Shulman~\cite{Shulman14} extensively explored dependencies in DNS and showed that an attacker can learn the requested domain in an encrypted DNS packet when information leakage via transitive trust is applied in tandem with other side-channels. Ateniese \etal~\cite{AtenieseM01} introduced a new strategy to build chains of trust from root servers to authoritative servers. End-to-end encryption has high overhead that needs to be mitigated. Besides, DNSSEC is not widely deployed yet even though DNS names are used for authentication. Thus, protections need to utilize encryption together with other methods such as DNSSEC, query name minimization, etc.

\BfPara{Modeling DNS Adversaries} While much of the previous work has focused on various aspects of DNS security and privacy, including (data-driven) modeling and informal description~\cite{MohaisenM15} and {\em informal adversaries modeling}   on confidentiality~\cite{BarnesSJHTTHB15}, there is no study that formalizes adversaries against confidentiality, let alone concretely evaluating them. 

\BfPara{Standards} The Internet Engineering Task Force (IETF) has recently established a working group dedicated solely to addressing DNS privacy concerns (DNS PRIVate Exchange, DPRIVE). This working group has proposed various techniques that are currently being under consideration~\cite{MohaisenM15}. Zhu \etal~\cite{HuZHMWH15a} (based on \cite{ZhuHHWMS15}) proposed a connection-oriented DNS transport over TCP, which uses TLS for privacy. The authors argue that the overhead of their approach is modest with careful implementations. Reddy \etal~\cite{ReddyWP15} proposed to use the Datagram Transport Layer Security (DTLS) for DNS exchange. They add a protection layer for the sensitive information in DNS queries, which would withstand a passive listener and certain active attackers. To address side-channel attacks on encrypted DNS, Mayrhofer~\cite{Mayrhofer15} proposed a padding scheme, where servers pad requests and responses  by a variable number of octets. DNS over TLS does not consider known attacks on TLS, such as man-in-the-middle and protocol downgrade. The use of simple padding schemes alone is not sufficient to mitigate traffic analysis attacks. However, padding will organize a part of more complex mitigations for traffic analysis attacks that are probably to be developed over time.

\ref{tab:lit} summarizes the work in the literature as categorized above, with example work, contributions, and open directions. 

\begin{table*}
\begin{center}
\caption{Summary of research directions in the literature, main contributions, and follow up.}\label{tab:lit}
{\scriptsize
\begin{tabular}{|p{1.3cm}|p{2.8cm}|p{1.9cm}|p{6.2cm}|p{3.5cm}|}
\hline
Area & Topic & Example work & Main contribution & Open direction \\
\hline
Ecosystem analysis & DNS resolver behavior & \cite{CallahanAR13, SchompRA16, BanseHF12, SchompCRA13, ShulmanW15, HaoFP11} & $\bullet$ DNS behavior tracking in a real-world setting $\bullet$ development an analytical model for client interactions $\bullet$ exploring the behavioral properties of domains using the DNS infrastructure associated with the domain and the DNS lookup patterns & Behavior under DNS changes, open resolvers\\
				&DNS blocking& \cite{ThomasLS14, ScaifeCT15, AppelbaumM15} & $\bullet$ NXD request patterns analysis, blocking special queries &How blocking affects privacy 		\\
\hline
DNS security &Vulnerability analysis& \cite{SchompCRA14, DagonPLL08, XuBSY13, JacksonBBSB09, SchompAR14, DagonAVJL08, ChenMP15} & $\bullet$ measurement of the Internet's vulnerability to DNS record injection attacks $\bullet$ analysis of rogue DNS servers to create malicious DNS resolution paths $\bullet$ analysis techniques for hiding malicious DNS activities $\bullet$ evaluation of the cost-effectiveness of mounting DNS rebinding attacks $\bullet$ removing shared DNS resolvers $\bullet$ making DNS queries more resistant to poisoning attacks & Best practices and incentives for fixing vulnerabilities \\ 
 &Malicious domain detection& \cite{AntonakakisPLVD11,AntonakakisPNVALD12,RahbariniaPA15,JonesFPWA16,HaoFP11,BilgeKKB11,LuoTZSLNM15,PerdisciCG12,DagonPLL08, YadavRRR10,AntonakakisPDLF10, GaoYCPGJD13, SzurdiKCSFK14} & $\bullet$ detecting malware-related domain names by passively monitoring DNS traffic $\bullet$ detecting unauthorized DNS root servers using endpoint-based measurements $\bullet$ detecting domain fluxes in DNS traffic by looking for patterns inherent to domain names $\bullet$ dynamic reputation system using passive DNS query data and zone features of domains $\bullet$ detecting unknown malicious domain using temporal correlation in DNS queries & Addressing evasion and stealthiness\\ 
  &Modeling adversaries& \cite{BauM10, HerzbergS13, GoldbergNPRVZ15} & $\bullet$ modeling the cryptographic operations in DNSSEC $\bullet$ discovering a forgery vulnerability & Mathematical modeling with standard adversaries\\ 
\hline
 DNS Privacy&Quantifying leakage& \cite{KoningsBSW13, KrishnanM10, ZhaoHS07, PaxsonCJRSSSTVW13, Castillo-PerezG09}	& $\bullet$ quantifying queries and device names leak plenty of information about users $\bullet$ analyzing the context obtained from the prefetching queries, privacy-preserving query scheme (Range Query) $\bullet$ exact setting representing a tradeoff between the scope of potential leakage versus the number of possible detections &Gap analysis of policy vs. reality\\ 
 			&Privacy-preserving designs& \cite{ZhaoHS07, Castillo-PerezG09, HerrmannFLF14, LuT10}	& $\bullet$ concealing the actual queries using noisy traffic $\bullet$ PIR techniques $\bullet$ PPDNS built on top of DHTs and computational PIR & Rigorous analysis and evaluation	\\ 
			&DNS encryption& \cite{HerrmannFLF14, ZhuHHWMS15, Shulman14, AtenieseM01} & $\bullet$ lightweight privacy-preserving name resolution service (EncDNS) $\bullet$ end-to-end encryption $\bullet$ T-DNS using TLS $\bullet$ chains of trust from root servers to authoritative servers &How encryption meets privacy (and how not)		\\ 
			&Modeling adversaries& \cite{MohaisenM15, BarnesSJHTTHB15}	& $\bullet$ data-driven modeling and informal description $\bullet$ modeling of adversaries against confidentiality & Modeling pervasive capabilities (no prior work)		\\ 
			&Standards&\cite{MohaisenM15, HuZHMWH15a, ZhuHHWMS15, ReddyWP15, Mayrhofer15} & $\bullet$ connection-oriented DNS transport over TCP $\bullet$ DTLS for DNS exchange $\bullet$ protection layer for the sensitive information in DNS queries $\bullet$ padding scheme &Rigorous analysis of security, privacy, and trust 		\\ 
\hline
\end{tabular}\vspace{-8mm}
}
\end{center}
\end{table*}

\section{Challenges and Directions}\label{sec:challenges}
Now we highlight various directions that pose challenges in research on DNS: we call for more research based on data, and data-driven analysis (\ref{sec:datadriven}), privacy as a plug-in (\ref{sec:privacyplugin}), modeling adversaries (\ref{sec:advmodel}), attack surface analyses (\ref{sec:attacksurf}), and addressing the open resolvers phenomena (\ref{sec:openres}). 

\subsection{More Data-driven Analysis for Security}\label{sec:datadriven}
There has been an abundance of work on DNS data analysis for security, such as DNS behavior tracking, encryption, blocking, query name minimization, DNSSEC, DNS over TLS, the DTLS for DNS exchange, etc. However, the DNS query traffic has been increasing and the behavior of DNS usage and the DNS ecosystem have been changing over time. DNS queries can represent plenty of information. For example, an attacker can build highly accurate profiles of what users do on the Internet by eavesdropping on query streams and ultimately breaching a user's privacy. Even more, some companies target individual users and build profiles  for them based on their browsing seen in DNS traffic. They assemble such profiles as part of their own commercial activities. Although efforts to prevent this leakage have been ongoing, many problems are still open. An understanding of the problem coupled with previous developments in DNS is necessary. In addition, many functions must be modified for the new DNS ecosystem and further research using DNS data must be done for security.
\subsection{Privacy as a Plug-in}\label{sec:privacyplugin}
With the increase in Internet usage, malicious invasion of privacy using the DNS operation is on the rise. Techniques such as query name minimization, DNSSEC, and DNS over TLS, etc. to solve these problems exist, but a solution that satisfies the requirements of privacy as a plug-in is lacking: such solution should not require major modifications nor interfere with the existing standards of DNS. For example, DNSSEC allows users to verify DNS responses are correct, but does not protect privacy. Encryption provided by TLS eliminates opportunities for eavesdropping, but it is unclear what notions of privacy it provides. DNSSEC and DNS over TLS are independent and compatible protocols, although each solving different problems. Thus, it is necessary to understand the notions of privacy under various security models, and realizing privacy as a plug-in to the DNS existing infrastructure by not requiring major modifications.
\subsection{Modeling Adversaries}\label{sec:advmodel}
Although previous studies have concentrated on DNS security and privacy, including data-driven modeling and the infrastructure and investments made by the DNS providers, little work is done on formalizing and understanding the notion of pervasive adversaries. These pervasive adversaries have been widely considered as a potential threat to the privacy and security of communication on the Internet. 

Modeling such adversaries would be the first challenge to improve DNS privacy. We should be able to view adversaries as either a passive adversary or an active adversary. A passive adversary does not interfere with the resolution and is interested in associating queries with a user or a set of users. He can eavesdrop on the links between the stub resolvers and recursive resolvers, and the links between the recursive resolvers and authoritative resolvers. An active adversary can control over a recursive resolver. For example, it is a result of the compromise of the software of that recursive resolver or due to being the adversary's recursive resolver such as an open rogue resolver. Formalizing the advantage of the adversaries would be the second challenge since the goal of the adversaries is to breach the privacy of users. It is meaningful to quantify their advantage in breaching the DNS privacy for addressing the privacy issue. An extending formalization of the capabilities of the adversaries using real-world DNS resolution topology and DNS query data would be another challenge. Some stub resolvers may generate more queries than others. Thus, if such information is known by an adversary, the adversary may use such distribution to associate a query with a user more often than with another. 

The last challenge is understanding the advantages of such adversaries in light of various ongoing activities in the DNS research community, which include: encryption, query name minimization and blocking for DNS privacy. For example, blocking mechanisms at either the browser, recursive, or authoritative server can improve DNS privacy--but it must be implemented at once in order to ensure privacy. In reality, it is almost impossible that all browsers and recursive resolvers on the Internet perform blocking at the same time. The diversity of browsers and recursive software (even by the same vendor) on the Internet today make it difficult to implement timely and synchronized blocking. Thus, it is reasonable to assume only a partial deployment of such recommendations. It should also be noted that while a user can maintain their privacy through DNS blocking, non-blocking users who share the same DNS infrastructure may be inadvertently affected.

\subsection{Attack Surface Analyses}\label{sec:attacksurf}
The attack surface of the DNS resolution system is the entire public Internet between the end user's connection and the public DNS service. The attack surface analysis is concerned with enumerating potential and confirmed vulnerabilities, the attacks those vulnerabilities can be used to launch, and the implications of those attacks. 

In the DNS resolution system, there are several potential attacks for disrupting the resolution operation. There has been a long history of attacks on the DNS ranging from Denial of Service (DoS) attacks to targeted attacks requiring specialized software. For example, an attacker can attack DNS resolvers by exploiting vulnerabilities such as buffer overflow attacks which make them misbehave or crash. Moreover, an attacker can actually modify DNS resolver configuration files and replace the name server IP addresses with malicious IP addresses to cause DoS attacks. These high-profile attacks have affected various commercial companies, software vendors, websites, content distribution services, and ISPs. 

DNS amplification attacks utilize DNS servers for performing bandwidth-consumption DoS attacks. An attacker can ``spoof" look-up requests to DNS servers to hide the source of the exploit and direct the response to the target. Essentially, the attacker turns a small DNS query into a much larger payload directed at the target network. 

With cache poisoning, an attacker can attempt to insert a fake address record for an Internet domain into the DNS. If the server accepts the fake record, the cache is ``poisoned'' and subsequent requests for the address of the domain are answered with the address of a server controlled by the attacker. For as long as the fake entry is cached by the server, all subscriber's browsers or e-mail servers will automatically go to the address provided by the compromised DNS server. DNS cache poisoning attacks do not require substantial bandwidth or processing, nor do they require sophisticated techniques.

Quantifying the attack surface of DNS is important in understanding and managing the DNS resolution system, thereby improving DNS privacy. It identifies critical pieces of the DNS system that need to be modified to withstand against security threats. As aforementioned, evaluating the advantages of various adversaries under blocking (at either the browser or recursive resolver) and examining the difference between the probabilities against the entities observed and controlled by adversaries at the blocking point would be the challenge. 

\subsection{Addressing Open Resolvers}\label{sec:openres}

While open resolvers provide various benefits, such as answering DNS requests from external sources for anything, they currently pose a significant threat to the stability and security of the Internet. Just recently, open resolvers have been utilized for launching amplification attacks, calling for initiating a systematic study on their population, use, and distribution, and raising the awareness on their potential roles. For example, the open resolver project~\cite{openres} reported 32 million open resolvers, 28 million of which pose a significant threat, as of October 2013.  However, little is done on understanding the role each of those millions of resolvers plays, whether they are open intentionally or accidentally, and other aspects of their behavior. 

One open problem today is to understand those resolvers by perhaps analyzing their role, and understanding how they contribute to the good and bad use of the DNS as a service. Some of the open questions that are worth exploring---which may shed light on the role each of those resolvers play---include, among others, the following. 1) How well-represented are the open resolvers in typical DNS resolution systems, e.g., in popular TLDs? 2) How persistent are open resolvers over time in both the DNS resolution and open resolver ecosystems? 3) Is there any correlation between the volumes of DNS queries generated by those resolvers in the DNS resolution system and their actual size in the open resolver ecosystem? 4) Do open resolvers ``lie'' about responses for queries initiated by other clients? 5) Are open resolvers consistent in answering various clients for the same type of query?

Other open questions concerning open resolvers could potentially be answered through a characterization of those resolvers, such as  geographical distribution, and persistence characterization over a longer period of time between consecutive scans, along with implications of the findings in the main questions mentioned earlier. Ultimately, findings systematically obtained through answering those questions could help a reputation system of the open resolvers ecosystem to guide benign users in their use of those resolvers.

\section{Concluding Remarks}\label{sec:concluding}
In this paper, we review various works on DNS ecosystem, security issues, and privacy concerns. We point out the open research problems related to DNS data-driven analysis, privacy as a plug-in, modeling adversaries, and attack surface analyses. We expect that these challenges and directions will continue to be useful for improving DNS security and privacy.

 \vspace{-1mm}



%
%


\end{document}